\documentclass[11pt]{article}
\usepackage{bbold}
\usepackage{amsmath}
\begin{document}
\date{}
\title{Von Neumann's Impossibility Proof: Mathematics in the Service of Rhetorics}
\author{Dennis Dieks \\History and Philosophy of Science\\
Utrecht University\\ d.dieks@uu.nl}
\maketitle
\begin{abstract}
According to what has become a standard history of quantum mechanics, in 1932 von Neumann persuaded the physics community that hidden variables are impossible as a matter of principle, after which leading proponents of the Copenhagen interpretation put the situation to good use by arguing that the completeness of quantum mechanics was undeniable. This state of affairs lasted, so the story continues, until Bell in 1966 exposed von Neumann's proof as obviously wrong. The realization that von Neumann's proof was fallacious then rehabilitated hidden variables and made serious foundational research possible again. It is often added in recent accounts that von Neumann's error had been spotted almost immediately by Grete Hermann, but that her discovery was of no effect due to the dominant Copenhagen \emph{Zeitgeist}.

We shall attempt to tell a story that is more historically accurate and less ideologically charged. Most importantly, von Neumann never claimed to have shown the impossibility of hidden variables \emph{tout court}, but argued that hidden-variable theories must possess a structure that deviates fundamentally from that of quantum mechanics. Both Hermann and Bell appear to have missed this point; moreover, both raised unjustified technical objections to the proof. Von Neumann's argument was basically that hidden-variables schemes must violate the ``quantum principle'' that physical quantities are to be represented by operators in a Hilbert space. As a consequence, hidden-variables schemes, though possible in principle, necessarily exhibit a certain kind of contextuality.

As we shall illustrate, early reactions to Bohm's theory are in agreement with this account. Leading physicists pointed out that Bohm's theory has the strange feature that pre-existing particle properties do not generally reveal themselves in measurements, in accordance with von Neumann's result.  They did not conclude that the ``impossible was done'' and that von Neumann had been shown wrong.

 \end{abstract}

\section{Introduction}\label{intro}

Von Neumann's book ``Mathematische Grundlagen der Quantenmechanik''  (Mathematical Foundations of Quantum Mechanics)\cite{VN1}, published in 1932, is widely acclaimed as a milestone in the history of quantum mechanics. It is a pioneering work that, among other accomplishments, first introduced Hilbert space techniques in the study of quantum theory, considered thermodynamic problems in terms of density matrices and what we now call von Neumann entropy, and clearly formulated the measurement problem. However, in addition to all well-deserved praise the book has also earned itself the reputation of containing a blunder that has been detrimental to the conceptual development of 20\textsuperscript{th} century physics, namely the infamous proof that it is impossible to add ``hidden variables'' to the formalism of quantum mechanics so that the theory becomes deterministic and more complete in its description of physical properties. This proof is now generally considered to be a flagrant fallacy; even such a patent one that it is mysterious how a mathematical genius like von Neumann could commit it.

In 1996 a second reprint of the 1932 book appeared, to which the mathematical physicist Rudolph Haag contributed a preface \cite{VN1}.\footnote{In this article I base myself on this second reprint of the German original, so references to page numbers in the \emph{Grundlagen} are to this edition. Translations from this book and other sources are mine, unless otherwise indicated.} In his preface Haag notes: ``one has characterized his [\emph{i.e.\@ von Neumann's}] proof of the impossibility of hidden variables as stupid [\emph{dumm}]''. He immediately adds apologetically: ``one should not forget, however, that this occurred in the context of significant original contributions by von Neumann, in this case the insight that mixed states are described by density matrices, which do not form a simplex''. This implicit acknowledgement of the foolishness of von Neumann's proof echoes countless similar comments in the literature of the past five decades, starting with John Bell's criticism \cite{bell66} that ``von Neumann's `very general and plausible postulates' are absurd''.  The reputation of von Neumann's proof has kept on declining since this first attack by Bell. Along the way down, Bell \cite{bell82} wrote in 1982: ``But in 1952 I saw the impossible [\emph{according to von Neumann}] done. It was in papers by David Bohm. But why then had Born not told me of this `pilot wave'? Why did von Neumann not consider it?''  And in 1988 Bell \cite{bell88} heightened the story by exclaiming: ``Yet the von Neumann proof, when you actually come to grips with it, falls apart in your hands!... It's not just flawed, it's silly... You may quote me on that: The proof of von Neumann is not merely false, it is foolish!''

A further perspective on the matter was opened up by Jammer \cite{jammer}, who in his 1974 book \emph{The Philosophy of Quantum Mechanics} drew attention to a forgotten publication by the German philosopher and physicist Grete Hermann. According to Jammer, in this publication from 1935 Hermann had already identified the same weak spot in von Neumann's proof as Bell would uncover in 1966. Hermann's work was definitively saved from oblivion and incorporated into what now has become a standard story by Mermin \cite{mermin}, who in 1993 wrote: ``A few years later Grete Hermann, 1935, pointed out a glaring deficiency in the argument, but she seems to have been entirely ignored. Everybody continued to cite the von Neumann proof. A third of a century passed before John Bell, 1966, rediscovered the fact that von Neumann's no-hidden-variables proof was based on an assumption that can only be described as silly---so silly, in fact, that one is led to wonder whether the proof was ever studied by either the students or those who appealed to it to rescue them from speculative adventures''.

The picture that emerges from these quotations, and from many similar passages that can be found in the literature, is quite suggestive. As Belinfante \cite {belinfante} put it: ``The truth, however, happens to be that for decades nobody spoke up against von Neumann's arguments, and that his conclusions were quoted by some as the gospel''. Seevinck \cite{seevinck} concludes in the same vein: ``eventually von Neumann's proof became considered holy---it was received as Biblical wisdom that one should not challenge''.  The suggestion clearly is that von Neumann's result was rhetorically used by the ruling class of Copenhagen physicists to ban the thought that quantum mechanics could be incomplete, with the aim of securing the hegemony of the Copenhagen interpretation. This might be called a program of indoctrination---which was only undermined when Bell showed that the emperor wore no clothes and publicly exposed the virtually patent fact that von Neumann's proof was incorrect.

Against this well-entrenched account, the present paper argues that the actual history was considerably less simple. As we shall show, Bell, Mermin and others probably misunderstood and certainly misrepresented von Neumann's proof, which is valid and sensible---in this assessment we elaborate on a recent paper by Jeffrey Bub \cite{bub}. Moreover, Bell and other commentators did not give a correct account of the conclusion von Neumann drew from his proof: von Neumann never claimed to have excluded \emph{all possible} hidden variables, but only those that could fit in a quantum kinematical framework. Further, we argue that Grete Hermann's work, although interesting both historically and philosophically, did not succeed in coming to grips with von Neumann's proof either. Interestingly, Hermann's technical objections are in their details rather different from Bell's although---as I shall argue---equally misdirected. Moreover, her argumentation was not of the kind that can be expected to have caught the attention of the physics community, and this will have been an important factor in the poor reception of her work.

Finally, we review some contemporary observations (from the nineteen-fifties) on how von Neumann's proof relates to the hidden-variables theories of Bohm and de Broglie. As will become clear, both major Copenhagenists and prominent hidden variables proponents understood what had actually been proved by von Neumann. If anything, it appears that a clear rhetorical use of von Neumann's result was first of all made by \emph{opponents} of Copenhagen, who used the alleged obvious fallacy in the proof to cast doubt on the scientific judgment of the Copenhagen circle (or clique), thus discrediting the Copenhagen interpretation itself by implication.

\section{The plan of von Neumann's book: The Quantum Paradigm}\label{book}

In the three-page Introduction to his 1932 book von Neumann unfolds his general plan: he intends not to address specific quantum mechanical problems, but is going to focus on the general structure and interpretation of the new theory, with particular attention to the role of probability. As he announces, he will demonstrate how the statistical formulas of quantum mechanics follow naturally from a small number of basic assumptions of a qualitative nature, so that they are not so counterintuitive as may appear at first sight.\footnote{As Lacki observes in \cite{lacki}, von Neumann's approach here can be seen as a continuation of Hilbert's program of axiomatizing physical theories.}  Further, he promises us to investigate in depth the issue of whether the statistical character of the theory can be understood as a manifestation of our lack of knowledge, in which case the existence of ``hidden variables'' would explain the need for probability statements. Concerning the latter issue, von Neumann already now reveals his main conclusion \cite[pp.\@ 2--3]{VN1}: ``It turns out however that this [\emph{i.e.\@ the introduction of hidden variables}] will hardly be possible in a satisfactory way---more precisely, such an explanation is incompatible with certain qualitative basic postulates of quantum mechanics''.\footnote{Indessen zeigt es sich, da{\ss} dies kaum in befriedigender Weise gelingen kann, genauer: eine solche Erkl\"{a}rung ist mit gewissen qualitativen Grundpostulaten der Quantenmechanik unvereinbar.}

Von Neumann's actual discussion of the hidden-variables question, leading to his (in)famous impossibility proof, only takes place in the fourth of the book's six chapters. In order to elucidate the status of the qualitative axioms on which that discussion is based we shall first briefly go through the book's first three chapters. But it should be noted already now that in his introductory remarks von Neumann does not declare that hidden variables are flatly impossible, but only that their introduction would conflict with what he considers basic general characteristics of quantum mechanics.

In chapter I of the \emph{Grundlagen}, von Neumann reviews the history of quantum mechanics leading up to matrix mechanics (Heisenberg, Born, Jordan) on the one hand and wave mechanics (Schr\"{o}dinger) on the other. Though both theories ultimately yield the same physical predictions, they start from very different principles and are also mathematically disparate. In particular, in matrix mechanics one has to operate with discrete quantities (components of matrices) whereas wave mechanics has the form of a continuum theory, so that a direct corrrespondence between the two structures appears impossible. Still, their physical equivalence suggests that both theories latch on to the same physical core.

The following similarity in the two approaches suggests a way of access to this essential core: both matrix and wave mechanics start from the classical description of physical problems in which a Hamiltonian is given as a function of coordinates $q_j$ and conjugated momenta $p_j$. In matrix mechanics these coordinates and momenta are replaced by matrices $Q_j$ and $P_j$, respectively, subject to the canonical commutation relation $P_j Q_j - Q_j P_j = \frac{\hbar}{i} \mathbb{1} $, after which the task becomes that of diagonalizing the resulting Hamiltonian matrix by finding its eigenvalues. In wave mechanics, on the other hand, one transforms the classical Hamiltonian into an operator working on continuous and differentiable functions $\psi$ of the $q_j$ (i.e., functions in classical configuration space) by replacing each occurrence in the Hamiltonian of $p_j$ by $\frac{\hbar}{i} \frac{\partial}{\partial q_j}$, after which one solves the differential equation $H \psi = \lambda \psi$. In both cases the solution of an eigenvalue problem is the central task, which suggests the existence of a physically meaningful isomorphism between the quantities that occur in these respective eigenvalue problems. So one expects an isomorphism that maps the matrix Hamiltonian to the Schr\"{o}dinger differential operator (and \emph{vice versa}), and the vectors in the matrix formalism to the functions $\psi$ in the Schr\"{o}dinger theory.

Although the discrete index space of matrix mechanics and the continuous classical configuration space of wave mechanics cannot be heaped together mathematically \cite[p.\@ 15]{VN1}, this thus proves to be an unimportant detail of mathematical surplus structure: the only things that matter are states (vectors and functions, respectively) and operators acting on them.\footnote{... die R\"{a}ume $Z$ und $\Omega$ sind wirklich sehr verschieden, und jeder Versuch, sie in Beziehung zu setzen mu{\ss} auf gro{\ss}e Schwierigkeiten sto{\ss}en. Das, worauf es uns in Wahrheit ankommt, ist gar nicht eine Beziehung von $Z$ zu $\Omega$, sondern nur eine solche zwischen ihren bzw.\@ Funktionen: d.h.\@ zwischen den Folgen $x_1, x_2, ..., $ die die Funktionen in $Z$ sind, und den Wellenfunktionen $\varphi(q_1... q_k)$, die die Funktionen in $\Omega$ sind. Denn diese sind es allein, die in die Fragestellungen der Quantenmechanik eingehen.}
Von Neumann draws the conclusion that wave mechanics and matrix mechanics are steppingstones to a pure quantum theory that is independent of  representational particularities and captures what is really physically significant. The corresponding mathematical structure is ``the abstract Hilbert space'', to which a detailed mathematical investigation is devoted in chapter II of the \emph{Grundlagen}.

This chapter II is the longest in the book and has the form of an independent mathematical treatise on the structure of Hilbert space. Notoriously, von Neumann here insists on mathematical rigour and avoids ``subterfuges'' like Dirac's delta-function. Although this sometimes complicates the presentation and notation, it can be said that Chapter II pioneers the complete formal machinery that has now become standard in quantum theory. In particular, attention is paid to the operator calculus in Hilbert space and to eigenvalue problems.

After the mathematical intermezzo of chapter II, chapter III (``The quantum mechanical statistics'') returns to the physical meaning of the formalism. In chapter I it had become clear that in the transition from classical theory to quantum mechanics the coordinates $q_j$ and momenta $p_j$ had to be replaced by either matrices or differential operators, and that these in turn should be seen as particular representations of  operators in an abstract Hilbert space. As a consequence, the classical Hamiltonian (representing the system's energy), being a function of the coordinates and momenta, was transformed into an operator as well. Von Neumann expands this into a general lesson: it is characteristic of quantum mechanics that physical quantities (i.e.\@ quantities that in principle can be measured on a physical system and characterize the system) are to be represented one-to-one by operators in Hilbert space \cite[sect.\@ III.1 and especially sect.\@ III.5]{VN1}.\footnote{These operators corresponding to physical quantities should be hypermaximal, i.e.\@ self-adjoint.}

Generalizing the proposals of Born and others concerning the physical meaning of the formalism \cite[III.1]{VN1}, von Neumann now derives the following rule: if $\textbf{R}$ is a physical quantity\footnote{Von Neumann systematically distinguishes between physical quantities and their representation in the mathematical formalism. Physical quantities (like energy, momentum, position) can be measured in the laboratory; they are denoted by bold-faced capitals like $\textbf{R}$. Their formal counterparts (operators in Hilbert space) are denoted by non-bold capitals like $R$.}, and $R$ the corresponding operator, the \emph{expectation value} of $\textbf{R}$ in the state $\varphi$ is given by $(R\varphi, \varphi)$, i.e.\@ the inner product in Hilbert space between $R\varphi$ and $\varphi$.\footnote{In the wave mechanics form of the theory the Hilbert space state $\varphi$ would correspond to a wavefunction $\varphi(q_1...q_k)$, in the matrix formulation it would be a vector $(x_1, x_2,...)$ on which the matrices act.} This formula for the expectation value of a physical quantity $\textbf{R}$ can be given the equivalent form
\begin{equation*}
  Exp(R) = \textbf{Tr}(P_{[\varphi]}R),
\end{equation*}
in which $P_{[\varphi]}$ is the projection operator on the state $\varphi$ and $\textbf{Tr}$ represents the trace (the result of taking the diagonal elements of an operator in its matrix representation).

This formula can be generalized for the case in which there is uncertainty about the quantum state \cite[pp.\@ 157--158]{VN1}. If there are various possibilities for the state, viz.\@ $\varphi_1, \varphi_2, ...$, with associated probabilities $w_1, w_2, ...$ quantifying our lack of knowledge, we obtain for the expectation value of $\textbf{R}$ the expression $\sum_i w_i (R\varphi_i, \varphi_i)$. This can alternatively be written as
\begin{equation}\label{trace}
 Exp(R) = \textbf{Tr} (U R),
\end{equation}
with $ U = \sum_i w_i P_{[\varphi_i]}$. The operator $U$ in this formula is Hermitian and non-negative, with trace equal to $1$. Since knowledge of $U$ suffices to determine the expectation values of all physical quantities in an ensemble, $U$ (the density operator) completely characterizes that ensemble. Equation \ref{trace} is the central statistical formula of quantum mechanics.

Via Equation \ref{trace}, and its alternate expressions, probabilities enter the theory, and it is natural to wonder about the status and origin of these probabilities. In particular, it should be made clear whether they play the same role as probabilities in classical statistical mechanics, that is, whether they quantify a lack of information about the micro-state. If so, this would imply the existence of ``hidden parameters'' that would be needed in addition to the quantum state to fix the actual physical state. Lack of knowledge of these hidden parameters would then explain the appearance of statistical arguments. As von Neumann observes, in classical physics the search for explanations in terms of such hidden parameters (or ``hidden variables'') has had many successes and has led to considerable progress, for example in the kinetic theory of gases \cite[p.\@ 109]{VN1}. He continues:
\begin{quote}
  Whether such an explanation by means of hidden variables is also appropriate in the case of quantum mechanics is an often asked question. The view that it will once be possible to answer it affirmatively has also now prominent representatives. Such an affirmative answer, when justified, would make the present form of the theory provisional, since in this case the $\varphi$-description of the states would be essentially incomplete.

  We shall demonstrate later (IV.2) that an introduction of hidden parameters is certainly not possible without changing the present theory in essential respects.\footnote{Ob f\"{u}r die Quantenmechanik eine derartige Erkl\"{a}rung durch verborgene Parameter in Frage kommt, ist eine viel er\"{o}rterte Frage. Die Ansicht, da{\ss} sie einmal in bejahendem Sinne zu beantworten sein wird, hat auch gegenw\"{a}rtig hervorragende Vertreter. Sie w\"{u}rde, wenn sie berechtigt w\"{a}re, die heutige Form der Theorie zu einem Provisorium stempeln, da dann die $\varphi$-Beschreibung der Zust\"{a}nde wesentlich unvollst\"{a}ndig w\"{a}re.
  Wir werden sp\"{a}ter (IV.2) zeigen, da{\ss} eine Einf\"{u}hrung von verborgenen Parameter gewi{\ss} nicht m\"{o}glich ist, ohne die gegenw\"{a}rtige Theorie wesentlich zu \"{a}ndern.}
\end{quote}
As in his earlier quoted statement (first paragraph of this section), von Neumann here anticipates what he is going to prove in chapter IV: he  announces that the structure of a hidden-variables theory must differ in an essential way from that of quantum mechanics. In the next section we shall see exactly what will have to be different in such a new theory.

Before we move on to chapter IV, we should pay attention to von Neumann's discussion of physical quantities in chapter III. In classical physics, measurable physical quantities were represented by functions on phase space; since Hilbert space should be regarded the successor of classical phase space, physical quantities will now have to correspond to operators in Hilbert space. But Hilbert space operators do not always commute, and this leads to complications when we try to apply the formulas for probabilities and expectation values to joint measurements of several quantities. Indeed, the formula $(R\varphi, \varphi)$ would not yield sensible results if we tried to represent by $\textbf{R}$ (i.e.\@ the physical quantity corresponding to the operator $R$) the simultaneous measurement of quantities represented by non-commuting operators. Von Neumann comments that this problem cannot be removed by some easy modification of the theory; the representation of physical quantities by non-commuting operators is a basic feature of the mathematical structure.\footnote{in der Struktur des mathematischen Werkzeugs der Theorie begr\"{u}ndet \cite[p.\@ 110]{VN1}} He proceeds to investigate its physical interpretation and concludes that quantities represented by commuting operators can be \emph{jointly} measured, whereas this is not possible for quantities corresponding to non-commuting operators.

An important corollary, emphasized by von Neumann, is that functions of operators like $R + S$ have a simple interpretation when $R$ and $S$ commute: in this case $R + S$ just represents the physical quantity $\textbf{R} + \textbf{S}$ that can be determined by measuring $\textbf{R}$ and $\textbf{S}$ jointly and adding the individual results. But in the case of non-commuting operators $R$ and $S$, this interpretation is not possible. Although $R + S$ will certainly represent \emph{some} measurable quantity of the quantum system, it is not at all clear what this quantity will be.\footnote{..., da{\ss} es gar nicht klar ist, was f\"{u}r nicht gleichzeitig me{\ss}bare $\textbf{R}$, $\textbf{S}$ unter $a\textbf{R} + b\textbf{S}$ verstanden werden soll \cite[p.\@ 131]{VN1}.} If $\textbf{R}$ and $\textbf{S}$ cannot be measured at the same time, the value of the quantity represented by $R + S$ need not have any relation to the values of $\textbf{R}$ and/or those of $\textbf{S}$, and the measuring procedure associated with $R + S$ must be expected to be independent of those appropriate for $\textbf{R}$ and $\textbf{S}$ separately.

In section III.5 of his book, ``Projection operators as propositions'', von Neumann further elaborates on his leitmotif that physical quantities are to be represented by Hermitian operators in Hilbert space. He now shifts his attention from quantities that are \emph{measured} on a system to \emph{properties} \emph{possessed} by the system, and considers yes-no propositions about such properties. This leads him to the result that such yes-no propositions, and therefore the physical properties of a quantum object, are represented by projection operators in Hilbert space. Consequently, a logical calculus of propositions can be founded on the structure of the lattice of projection operators in Hilbert space.\footnote{Wie man sieht, erm\"{o}glicht die Beziehung zwischen den Eigenschaften eines physikalischen Systems einerseits und den Projektionsoperatoren andererseits eine Art Logikkalk\"{u}l mit diesem \cite[p.\@ 134]{VN1}.} Of course, the projection operators do not form a commutative algebra, so that a non-classical ``quantum logic'' emerges in this way, something to be famously explored in more depth later by von Neumann together with Birkhoff \cite{birkhoff}.

Summing up, the first three chapters of von Neumann's book distill an essential core out of the two different historical forms of the theory that had developed in the nineteen-twenties. This ``quantum theory proper'' establishes a new kinematical standard for physics: it is characteristic of quantum physics that physical quantities and properties are represented by operators in Hilbert space. This new ``quantum paradigm'', quantum kinematics, will play an important role in von Neumann's project, in chapter IV, of deducing the statistical features of quantum mechanics from first principles.

\section{The Impossibility Proof}\label{proof}

In chapter IV, ``Building up the theory deductively'' (Deduktiver Aufbau der Theorie), first section ``Fundamental justification of the statistical theory'' (Prinzipielle Begr\"{u}ndung der statistischen Theorie), von Neumann sets himself the task of deriving the basic formula for expectation values, Equation \ref{trace}, from a small number of general and qualitative assumptions that are independent of the details of quantum mechanics. As he proudly announces, in the process it will become possible to verify the whole statistical framework of quantum mechanics, as laid out in chapter III.\footnote{This independent deduction of the statistical core of quantum mechanics can already been found in \cite{VN2}. However, in this older publication there is no emphasis on the hidden-variables question.}

To set the stage, von Neumann asks us \emph{to forget everything about quantum mechanics} and to consider physical systems as experimentally defined by measurable quantities \cite[p.\@ 158]{VN1}. In the case of jointly measurable quantities $ \textbf{R}, \textbf{S},...$ we can immediately include arbitrary functions of $\textbf{R}$ and $\textbf{S}$ in the collection of physical quantities. Indeed, such functions $f(\textbf{R},\textbf{S})$ can simply be defined by applying the function $f$ to the measurement results for $\textbf{R}$ and $\textbf{S}$ (measured jointly).

However, von Neumann continues, one should be aware that the attempt to similarly form $f(\textbf{R},\textbf{S})$ for a single system is senseless in the case that $\textbf{R}$ and $\textbf{S}$ are not jointly measurable: in this situation it is completely unclear what the associated measuring procedure should be.\footnote{Man vergegenw\"{a}rtige sich aber, da{\ss} es vollkommen unsinnig ist, $f(\textbf{R},\textbf{S})$ bilden zu wollen, wenn $\textbf{R}$, $\textbf{S}$ nicht gleichzeitig me{\ss}bar sind: es gibt ja keinen Weg, die dazugeh\"{o}rigen Me{\ss}anordnung anzugeben \cite[p.\@ 158]{VN1}.}

However, we need not restrict ourselves to the consideration of single systems, but can also consider statistical collectives of systems. In such collectives, we may measure $\textbf{R}$ on one sub-collective, and $\textbf{S}$ on another sub-collective. This makes it possible to \emph{define} a physical quantity like $\textbf{R} + \textbf{S}$, \emph{even if $\textbf{R}$ and $\textbf{S}$ are not jointly measurable}, namely as a quantity that satisfies the additivity condition on expectation values in the collective: $Exp(\textbf{R}+\textbf{S}) = Exp(\textbf{R}) + Exp(\textbf{S})$ \cite[p.\@ 164]{VN1},\cite{bub}.

Von Neumann emphasizes once again at this point that in the case of failing co-measurability the quantity $\textbf{R}+\textbf{S}$, if it could be measured directly, must be assumed to be operationally independent of $\textbf{R}$ and $\textbf{S}$. The definition via expectation values defines $\textbf{R} + \textbf{S}$ only indirectly, without giving any indication whether, and if so how, a direct measuring procedure for it will relate to measuring procedures for $\textbf{R}$ and $\textbf{S}$.\footnote{Wenn R und S gleichzeitig me{\ss}bar sind, mu{\ss} $\textbf{R} + \textbf{S}$ die gew\"{o}hnliche Summe sein. Im allgemeinen ist es aber nur auf implizite Weise gekennzeichnet, und wir k\"{o}nnen die Me{\ss}vorschriften f\"{u}r $\textbf{R}$, $\textbf{S}$ kaum zu einer solchen f\"{u}r $\textbf{R} + \textbf{S}$ zusammensetzen \cite[p.\@ 164]{VN1}.}

In a note (note 164) von Neumann illustrates the situation with the example of the quantum operator that represents the energy of an electron in a potential field: $H = f(Q) + g(P)$. A measurement of $\textbf{H}$ may be performed via the determination of spectral frequencies, whereas the position $\textbf{Q}$ and the momentum $\textbf{P}$ will be measured in completely different and unrelated ways---moreover, the three quantities cannot be measured jointly. Still, we have $Exp(\textbf{H}) = Exp\{f(\textbf{Q})\} + Exp\{g(\textbf{P})\}$. It should be noted, however, that although this illustration comes from quantum mechanics, von Neumann's above implicit definition of $\textbf{R} + \textbf{S}$ by means of statistical collectives is independent of quantum mechanics and only makes use of operationally defined notions.

The upshot of the above is that once we can operationally define physical quantities for a system, we can also consider sums of these quantities, whether they are co-measurable or not. In the case of co-measurable quantities the characterization of the sum quantity is trivial: we just add the values of the quantities that we sum. In the case of quantities that cannot be measured together we only define the sum implicitly: $\textbf{R} + \textbf{S}$ is a certain quantity whose expectation values in an arbitrary statistical collective equals the sum of the expectation values of $\textbf{R}$ and $\textbf{S}$, respectively.

The argument leading to von Neumann's main result is now relatively straight-forward. As von Neumann states, we have seen in the earlier chapters that the essential ingredient of the ``quantum transition'' is the representation of physical quantities by Hermitian operators in Hilbert space, keeping functional relations between the quantities intact. He therefore assumes that if we are going to construct a quantum theory, we should make each measurable physical quantity $\textbf{R}$ correspond to an operator $R$. Moreover, $f(\textbf{R})$ will have to correspond to the operator $f(R)$; and the quantity $\textbf{R} + \textbf{S}$ to the operator $R + S$. In the latter case $\textbf{R} + \textbf{S}$ stands for the implicitly defined quantity introduced above if $\textbf{R}$ and $\textbf{S}$ are not jointly measurable.\footnote{Von Neumann makes the further uncontroversial assumption that the expectation value of any non-negative quantity must be non-negative.}

Remarkably, these definitions and assumptions are strong enough to obtain \cite[pp.\@ 167--168]{VN1} the far-reaching conclusion that in any thus constructed theory the expectation value of an arbitrary physical quantity $\textbf{R}$, in an arbitrary statistical ensemble of physical systems, can be written as $Exp(R) = \textbf{Tr}(UR)$. Here $U$ is a non-negative unity-trace Hermitian operator that characterizes the ensemble (so $U$ is independent of $R$). In other words, von Neumann has demonstrated that any possible statistical collective can be characterized by a density operator $U$, and that all expectation values possess the form of Equation \ref{trace}. Importantly, he has not derived this by presupposing the full framework of quantum mechanics. He has based himself on more general assumptions: \textbf{I}, that the operationally defined physical quantities\footnote{Partly implicitly defined!} of any physical system correspond one-to-one to to Hermitian operators in Hilbert space and, \textbf{II}, that this correspondence respects addition relations.

From the general validity of the expression $Exp(R) = \textbf{Tr}(UR)$ it follows as a corollary that there are no completely dispersion-free ensembles for quantum quantities: whatever choice we make for the operator $U$, it is always possible to find operators $R$ in Hilbert space such that the statistical spread of $R$, calculated with $U$ via the trace formula, does not vanish.\footnote{I.e., $Exp(R^2) - \{Exp(R)\}^2 > 0$.} This excludes hidden variables: indeed, the very idea of the introduction of such variables is to have the possibility of ensembles \emph{without} statistical spreads, namely ensembles in which the hidden parameters possess fixed values. If such parameters existed, the spreads actually predicted by quantum mechanics would have to result from averaging over dispersion-free ensembles, corresponding to sub-quantum dispersion-free states. But von Neumann's proof shows that, given his premises, there \emph{are no} such finer-grained states.

Von Neumann himself summarizes the result of his no-hidden-variables proof as follows \cite[p.\@ 171]{VN1}:
\begin{quote}
 ``it is impossible that the same physical quantities, with the same mutual relations, are present (i.e.\@ that our premises \textbf{I} and \textbf{II} hold), if in addition to the wave function yet other variables (``hidden parameters'') exist.

 It would not help if in addition to the familiar quantities that are represented by operators in quantum mechanics new, still undiscovered quantities existed: for already in the case of the familiar quantities the quantum mechanical relations (i.e.\@ \textbf{I}, \textbf{II}) must fail. It is therefore not, as is often assumed, a question of interpretation of quantum mechanics---the system of quantum mechanics would have to be objectively false in order that another description of the elementary process than the statistical one be possible.\footnote{...ja es ist sogar ausgeschlossen, da{\ss} dieselben physikalischen Gr\"{o}{\ss}en mit denselben Verkn\"{u}pfungen vorhanden sind (d.i.\@ da{\ss} \textbf{I}, \textbf{II} gelten), wenn neben der Wellenfunktion noch andere Bestimmungsst\"{u}cke (``verborgene Parameter'') existieren sollen. Es w\"{u}rde nicht gen\"{u}gen, wenn au{\ss}er den bekannten, in der Quantenmechanik durch Operatoren repr\"{a}sentierten, physikalischen Gr\"{o}{\ss}en noch weitere, bisher unentdeckte, existierten: denn schon bei den erstgenannten, bekannten Gr\"{o}{\ss}en m\"{u}{\ss}ten die von der Quantenmechanik angenommenen Verkn\"{u}pfungen (d.i.\@ \textbf{I}, \textbf{II}) versagen. Es handelt sich also gar nicht, wie vielfach angenommen wird, um eine Interpretationsfrage der Quantenmechanik, vielmehr m\"{u}{\ss}te dieselbe objektiv falsch sein, damit ein anderes Verhalten der Elementarprozesse als das statistische m\"{o}glich wird.}
\end{quote}
Here, again, \textbf{I} and \textbf{II} refer to the assumptions that each physical quantity corresponds to a Hermitian operator in Hilbert space, and that sums of such quantities correspond to the sums of the corresponding operators, respectively. In other words, the proof tells us that the hidden physical properties added to the quantum description in a hidden variables completion of the theory cannot correspond to Hermitian operators in Hilbert space in the way the standard quantum quantities do. If we were to add finer-grained values to the usual quantum quantities (represented by operators), these values would consequently have to violate the functional relations between the operators (in general, they could not accord with \textbf{II}). For example, if we assigned the hidden value $r$ to the quantity represented by $R$ in quantum mechanics, the hidden value $s$ to $S$, and the hidden value $t$ to $R + S$, it would generally not be possible to have $t = r + s$.

Does von Neumann's proof show that no hidden-variables theories can exist at all? Clearly not.\footnote{One should perhaps add that it is difficult to make sense at all of the claim that certain sets of empirical data are impossible to imbed in any deterministic model whatsoever. As Bertrand Russell argued in a famous essay of 1913 \cite{russell}, determinism all by itself is empirically empty. The doctrine of determinism only obtains empirical bite if one specifies in some detail what the deterministic theory should look like. \label{Noteunderdet}} If one grants the possibility that quantum mechanics may be superseded by a theory that can \emph{not} be formulated in Hilbert space, von Neumann's theorem loses its cogency. Von Neumann has the following to say about this possibility \cite[p.\@ 173]{VN1}:
\begin{quote}
  in the present state of our knowledge everything speaks against this: for the only presently available formal theory that orders and summarizes our experiences in a somewhat satisfactory way, namely quantum mechanics, is in strict logical conflict with it. However, it would be an exaggeration to claim that this is the absolute end of causality: there are certainly gaps in quantum mechanics in its present form, and it may even be that it is false, although the latter is quite improbable in light of its astounding achievements in understanding general problems and in making specific calculations [....] one can never say of a theory that it has been proved by experience [...] But even considering all these words of caution we are allowed to say: at this moment there is no reason and no excuse to keep on talking about causality in nature---for no experience supports its existence, as macroscopic experience is unable to do so as a matter of principle, and the only known theory that is compatible with our experiences concerning elementary processes, quantum mechanics, contradicts it.\footnote{Hier spricht aber beim heutigen Stande unserer Kenntnisse alles dagegen: denn die einzige zur Zeit vorhandene formale Theorie, die unsere Erfahrungen in halbwegs befriedigender Weise ordnet und zusammenfa{\ss}t, das ist die Quantenmechanik, steht mit ihr in zwingendem logischen Widerspruch. Es w\"{a}re freilich eine \"{U}bertreibung, zu behaupten, da{\ss} die Kausalit\"{a}t damit abgetan ist: die Quantenmechanik ist in ihrer heutigen Form gewi{\ss} l\"{u}ckenhaft, und es mag sogar sein, da{\ss} sie falsch ist, wenngleich dies letztere angesichts ihrer verbl\"{u}ffenden Leistungsf\"{a}higkeit beim Verst\"{a}ndnis allgemeiner und der Berechnung spezieller Probleme recht unwahrscheinlich ist. [...] kann man doch niemals von einer Theorie sagen, sie sei durch die Erfahrung bewiesen [...] Aber bei Beachtung aller dieser Kautelen d\"{u}rfen wir doch sagen: es gibt gegenw\"{a}rtig keinen Anla{\ss} und keine Entschuldigung daf\"{u}r, von der Kausalit\"{a}t in der Natur zu reden---denn keine Erfahrung st\"{u}tzt ihr Vorhandensein, da die makroskopischen dazu prinzipiell ungeeignet sind, und die einzige bekannte Theorie, die mit unseren Erfahrungen \"{u}ber die Elementarprozesse vertr\"{a}glich ist, die Quantenmechanik, widerspricht ihr. \label{notevN}}
\end{quote}
So von Neumann's position is that quantum mechanics, given its history of unexpected and astounding achievements, is so well supported that there is no reason to doubt the validity of its core theoretical structure. And from the acceptance of this Hilbert space structure it follows compellingly that ``causality'' is impossible to implement, in the sense that dispersion-free states cannot be accommodated.

Summing up, in his chapter IV von Neumann has proved that viable hidden-variables theory cannot be Hilbert space theories; such theories must violate the ``quantum paradigm''. Therefore, hidden values of at least some physical quantities will not obey the same relations as the corresponding quantum observables: as already pointed out above, if two such quantum observables add up to a third one, their hidden values will generally not add up in the same way.
Conversely, the physical quantities that were implicitly defined by von Neumann will in a hidden-variables theory generally not correspond to measurement results. To see this, note that the hidden value of the quantity $\textbf{R} + \textbf{S}$ will be given by the sum of the values of $\textbf{R}$ and $\textbf{S}$, respectively, also when $R$ and $S$ do not commute. If we accept the empirical adequacy of quantum mechanics on the statistical level, this sum value will not be a measurement result (as measurement results are eigenvalues of operators). So experimental outcomes will not reflect all the properties that are actually present in the system, according to the hidden-variable theory: measurements will not completely represent the system as it is in itself. This introduces an element of ``measurement contextuality'' in hidden-variables theories.

\section{Comparison with Bell's criticism}\label{Bell}

In the Introduction of his 1966 paper ``On the Problem of Hidden Variables in Quantum Mechanics'' \cite{bell66} John Bell asserts that von Neumann's hidden-variables argument ``leaves the real question untouched''. In section III of the same paper Bell summarizes von Neumann's argument, stating that von Neumann's essential assumption is the following: ``\emph{Any real combination of any two Hermitian operators represents an observable, and the same linear combination of expectation values is the expectation value of the combination}'' \cite[pp.\@ 448--449]{bell66}. Bell notes that in the case of dispersion-free states the expectation value is equal to the fixed value of the physical quantity in question, and if quantum mechanics is empirically adequate this value should be an eigenvalue of the operator representing the quantity. But then it can be seen immediately that the above assumption cannot be satisfied: the eigenvalue of $R + S$ is not equal to the eigenvalue of $S$ plus the eigenvalue of $R$ if $S$ and $R$ do not commute---eigenvalues do not combine linearly. Von Neumann's essential assumption is therefore evidently incorrect. The explanation of its failure is in fact only to be expected: ``A measurement of a sum of non-commuting observables cannot be made by combining trivially the results of separate observations on the two terms---it requires a quite distinct experiment''. To bring this point home Bell gives the example of spin measurements on a spin-$1/2$ particle in the $x$ and $y$ directions; a measurement of $\sigma_x + \sigma_y $ would in this case be a measurement in a third orientation, which requires a completely different and independent experimental set-up.

Therefore, Bell concludes, von Neumann's assumption is utterly inappropriate for dispersion-free states. Consequently, von Neumann's proof fails to demonstrate what von Neumann himself thought that it demonstrated, which Bell quotes (from the English translation by Beyer \cite{VN1}) as: ``It is therefore not, as is often assumed, a question of reinterpretation of quantum mechanics---the present system of quantum mechanics would have to be objectively false in order that another description of the elementary process than the statistical one be possible''.\footnote{As we shall discuss below, it is very important to place this statement in its proper context. In fact, it is \emph{the last sentence} of von Neumann's conclusion, which we have quoted in full in the previous section. \label{Note}} Bell interprets this as the statement that the predictions of quantum mechanics must be wrong if hidden variables exist, and objects that von Neumann has not proved this at all. As he points out, it was \emph{not} the objective measurable predictions of quantum mechanics that ruled out hidden variables, but it was rather von Neumann's own ``arbitrary assumption of a particular (and impossible) relation between the results of incompatible measurements either of which \emph{might} be made on a given occasion but only one of which can in fact be made'' \cite[p.\@ 449]{bell66}.

In his paper from 1982 \cite{bell82} Bell presents his criticism of von Neumann in a more compact form, as follows. He identifies the vital ingredient of the proof as the assumption that for linearly combined operators the measurement results are similarly linearly related. But this cannot possibly hold, for the individual results are eigenvalues, and as an elementary mathematical fact eigenvalues of linearly related operators are not linearly related themselves. So von Neumann's ``very general and plausible postulate'' is actually absurd \cite[p.\@ 994]{bell82}.

As we have seen in the Introduction, in 1988 Bell winds up by characterizing the proof as flawed and moreover silly \cite{bell88}.

Bell thus makes two claims: 1. von Neumann's proof is flawed, and 2. the premises of the proof are inappropriate. Concerning 1, Bell clearly does not mean that von Neumann made a mathematical error. Rather, the claim is that von Neumann drew a wrong conclusion; he overstated the significance of his proof. According to Bell, von Neumann thought that he had proved that the empirical predictions of quantum mechanics must be wrong if hidden variables exist---which cannot be right, as demonstrated, for example, by Bohm's theory. The evidence Bell adduces is von Neuman's final concluding statement (already quoted above): ``the present system of quantum mechanics would have to be objectively false in order that another description of the elementary process than the statistical one be possible.''

But as already noted, this sentence is merely the final part of von Neumann's comment on the significance of his proof (the complete quotation is in the previous section, see note \ref{notevN} and its translation in the main text). Von Neumann starts by saying that hidden variables will not be representable in the same way as quantities in standard quantum mechanics: assumptions \textbf{I} and \textbf{II} will have to be violated. Hidden variables can therefore not correspond to operators in Hilbert space. It is in \emph{this} sense that the system of quantum mechanics would be objectively wrong if it were to prove empirically unavoidable to introduce hidden variables: a hidden variables theory must necessarily jettison the core theoretical structure of quantum mechanics (see section \ref{proof}).

We can therefore conclude that the first arrow of Bell's criticism, that the proof is mistaken, is misdirected. Bell has misrepresented and almost certainly misunderstood what von Neumann argued for (some more about this in section \ref{conclusion}).

In the second part of his criticism Bell questions von Neumann's assumptions. Apparently, the idea here is that von Neumann's proof proves \emph{something}, even if this is different from what von Neumann assumed it proved: evidently, the proof demonstrates exactly what  follows from its premises. But, Bell argues, we should not be impressed by this objective content, as the premises are irrelevant. This is illustrated by the already-mentioned argument about eigenvalues: for dispersion-free states with hidden-variables values that equal the eigenvalues of non-commuting observables it is self-explanatory that von Neumann's premises cannot hold.

A remarkable aspect of this criticism is that it does not identify a mistake in von Neumann's argument. Indeed, as shown by von Neumann and discussed in section \ref{proof}, his assumptions exclude hidden-variable theories in which quantities that are represented in quantum mechanics by operators $R$, $S$ and $R+S$, possess expectation values (and therefore in the case of dispersion-free ensembles simply \emph{values}) related as $r$, $s$ and $r+s$, respectively. Remember that von Neumann was able to prove this quite generally---in particular, he did not assume that the hidden variables equal the eigenvalues of quantum operators. Now Bell objects, via an example, that if we want to construct a hidden-variable theory in which the quantities represented in quantum mechanics by operators $R$, $S$ and $R + S$ possess hidden values equaling the respective eigenvalues, this generally cannot be done if we require that the value of $R + S$ must equal $r + s$. But this just confirms by one particular illustration  what von Neumann proves in general.

To be fair, it should be acknowledged that Bell's example is not meant as a counterexample to von Neumann's assumptions, but rather as a rhetorical device that shows their irrelevance: one should not seriously expect von Neumann's premises to hold for any viable hidden-variables scheme anyway. Indeed, so the argument apparently goes, if one knows even only a bit about quantum mechanics, it is obvious that hidden-variables theories satisfying von Neumann's assumptions cannot exist. So the right conclusion of the proof (instead of von Neumann's own erroneous interpretation of it---according to Bell, that is) is trivial and does not add anything to what we can see immediately.

At the basis of this triviality argument is Bell's quick summary of von Neumann's premises in the form: ``\emph{Any real combination of any two Hermitian operators represents an observable, and the same linear combination of expectation values is the expectation value of the combination}'' \cite[pp.\@ 448--449]{bell66}. But here Bell misses a vital part of von Neumann's reasoning. As we have seen in section \ref{proof}, von Neumann starts his argument by discussing \emph{physical quantities} independently of quantum theory and independently of their mathematical representation. He \emph{defines} linear combinations of these quantities such that their expectation values \emph{per definition} satisfy the same linear relations as the quantities themselves (cf.\@ \cite{bub}). So for these physical quantities the statement that expectation values preserve linear relations is not an assumption, but an analytic truth. Now, differently from what Bell suggests in his 1966 paper, von Neumann takes into account that quantities may fail to be co-measurable and is aware that this casts doubt on the meaning of sums of such quantities---as we have seen in section \ref{proof}, he draws attention to this complication repeatedly. This difficulty is precisely the reason why he defines the sum of two non-jointly measurable quantities in an implicit way, thus securing the additivity of their expectation values. So it is false to claim that linear relations between physical quantities, as defined by von Neumann, cannot hold generally also between their expectation values and measurement values.

What von Neumann does assume (\emph{this} is his real assumption) is that the thus-defined linearly related physical quantities should be bijectively associated with linear operators in a Hilbert space; this assumption is based on his notion that classical kinematics should be replaced by a quantum kinematics of operators. From this assumption follows the standard quantum mechanical expression for expectation values (equation \ref{trace}) and the non-existence of dispersion-free ensembles. But it also follows that if there \emph{are} dispersion-free ensembles, in some hidden-variables scheme, \emph{not} all physical quantities of the system---as defined by von Neumann---can correspond to operators in Hilbert space. Because Bell overlooks von Neumann's independent definition of physical quantities, he misses this part of the argument.

For the case of Bell's spin example this means that the von Neumann physical quantity defined as $\textbf{s}_x + \textbf{s}_y$, i.e.\@ the quantity that in a hidden-variables scheme possesses the value that is simply the sum of the values of $\textbf{s}_x$ and $\textbf{s}_y$, cannot be represented by the operator $\sigma_x + \sigma_y$. The value of $\textbf{s}_x + \textbf{s}_y$ must remain hidden in measurements (assuming the empirical adequacy of quantum mechanics, so that all measurement results are eigenvalues of spin operators).

Bell thus misconstrues the premises of von Neumann's proof. He incorrectly interprets them as a requirement imposed on the expectation values of physical quantities that are defined via their representation by operators in standard quantum mechanics. His triviality objection boils down to the observation that it is easy to see that this requirement cannot be imposed anyway. By contrast, there is nothing obviously wrong or impossible when we follow von Neumann's own reasoning. If von Neumann's premises are formulated as von Neumann himself stated them, there is no triviality in his proof.

\section{Grete Hermann's criticism of von Neumann}\label{Hermann}

In 1933, a year after the publication of von Neumann's book, Grete Hermann wrote a paper ``\emph{Determinism and Quantum Mechanics}'' (``\emph{Determinismus und Quantenmechanik}'') \cite{hermann3}.\footnote{The paper remained unknown and unpublished until very recently. Hermann sent a copy to Dirac, in whose archive it was retrieved \cite[Ch.\@ 8]{crull}. An English translation has now been made available in \cite[Ch.\@ 14]{crull}.}
In this paper Hermann addresses the question of whether quantum mechanics can be said to disprove determinism. Her interest in this question comes from her sympathy for neo-Kantianism, according to which the ``law of causality'' must be \emph{a priori} valid if scientific knowledge is to be possible. Accordingly, she sets out to show that even if quantum mechanics were to prove completely right also in the future, this still would not constitute a refutation of the law of causality. Not surprisingly then, she takes exception to von Neumann's argument that the formalism of quantum mechanics contradicts a perfectly causal description of natural processes and devotes a section of her paper to a critical discussion of the 1932 proof.

In the relevant part of her paper, after mentioning von Neumann's assumptions \textbf{I} and \textbf{II}, Hermann comes to the now familiar point of the definition of sums of quantities that are not jointly measurable and the calculation of their expectation values. She writes \cite[p.\@ 250]{crull}:
\begin{quote}
However, this definition [\emph{i.e.\@ taking the ordinary sum}] fails for quantum mechanical quantities that have non-commuting
operators, because these quantities are not `simultaneously measurable' on one and the same physical state. The sum $\textbf{R} + \textbf{S}$ in this case can be defined only indirectly as the quantity corresponding to the sum $R + S$ of the operators belonging to $\textbf{R}$ and $\textbf{S}$. As Neumann shows in an instructive example, the eigenvalues of $R + S$ in no way need to be the sums of those of $R$ and $S$;
this, however, would be necessary to ensure that the proof of condition b [\emph{i.e.\@ the `condition' that linear combinations of quantities have expectation values that are the same linear combinations}] for expectation value functions could be carried over from the classical theory. Neumann thus needs another proof for quantum mechanics. He finds it in the following consideration: in the formalism of quantum mechanics, the expectation value of a quantity $\textbf{R}$ in the state $[\varphi]$ is given by the symbol $(R\varphi,\varphi)$, where $R$ is the operator belonging to $\textbf{R}$. Since $((R + S)\varphi, \varphi) = (R\varphi, \varphi)+ (S\varphi, \varphi)$ holds for this symbol, the expectation value of a sum is indeed equal to the sum of the expectation values.
\end{quote}

She then comments that the latter consideration is only convincing if one already \emph{assumes} that the quantum state is the only determining factor in the calculation of expectation values; that is, that as yet to be discovered new traits (i.e.\@ hidden variables) can play no role. As she says: ``Indeed, for ensembles of physical systems agreeing with one another besides in the wave function also in terms of such a newly discovered trait, it has not been shown that the expectation value function has the form $(R\varphi, \varphi)$ \cite[p.\@ 251]{crull}''.
Hermann concludes:
\begin{quote}
Therefore, in terms of the predictability of measurement results we have the following: as Neumann's proof shows, a physicist who only knows a given system by its Schr\"{o}dinger function [\emph{i.e.\@ $\varphi$}] is bound to limits conforming to the Heisenberg relations in predicting measurement results. More is not proven.

Whoever wants to extract more from Neumann's proof must already assume that $(R\varphi, \varphi)$ represents the average value for the eigenvalue measurements of $\textbf{R}$ for any ensemble whose elements, besides with respect to $\varphi$, agree with one another also with respect to arbitrary further conditions. But that all these ensembles have the same average values is an assumption justified neither by previous experience nor by the hitherto confirmed theory of quantum mechanics. Without it, the proof of indeterminism collapses.
\end{quote}

Hermann thus accuses von Neumann of a \emph{petitio} when he arrives at the conclusion that determinism is not tenable: she claims that he has assumed the irrelevance of hidden variables for expectation values to start with, so that it is no wonder that in his analysis they cannot help to make predictions more accurate.

At first sight, Hermann's criticism may seem virtually identical to Bell's objection, as it also centres on the question of whether expectation values may be added in the case of sums of quantities that are not co-measurable.\footnote{Jammer comments \cite[p.\@ 274]{jammer}: ``It is remarkable that Hermann's criticism touched precisely on one of the weakest points in the proof, the additivity postulate.'' Seevinck \cite[p.\@ 122]{seevinck}: ``Hermann concluded, as would Bell, that von Neumann ruled out the existence of dispersion-free states by requiring without further justification the additivity rule also at the level of hidden variables.'' As we have seen, Mermin \cite{mermin} contends that John Bell rediscovered Grete Hermann's objection.} But on closer inspection there is a major difference. As discussed in section \ref{Bell}, Bell suggested that von Neumann had not realized that eigenvalues fail to combine linearly and had therefore been seduced into accepting a postulate (the additivity postulate for expectation values) that he would never have taken seriously otherwise. By contrast, Hermann notes that von Neumann himself shows, via ``an instructive example'', that eigenvalues of $R + S$ in no way need to be sums of eigenvalues of $R$ and $S$.\footnote{Wie Neumann an einem aufschlussreichen Beispiel zeigt, brauchen die Eigenwerte von $R + S$ sich keineswegs aus denen von $R$ und $S$ zusammensetzen zu lassen \cite[p.\@ 19]{hermann3}. The example referred to here must be the one of note 164 in the \emph{Grundlagen}, in which von Neumann comments that the total energy, which is the sum of kinetic and potential energy, cannot be computed from these two components separately. In his 1927 paper \cite{VN2}, which already contains the no-hidden-variables proof, there is a similar example.} In direct opposition to Bell's reasoning, Hermann contends that it was his very awareness of this fact that made it clear to von Neumann that he needed a \emph{proof} of the additivity of expectation values for the case of quantities that are not simultaneously measurable. According to Hermann, for this proof von Neumann resorted to the fact that in quantum mechanics additivity holds generally, even in the case of non-commuting operators. This then leads to a \emph{petitio principii}, says Hermann: features that are unique to quantum mechanics are picked out to argue that only quantum mechanics itself can possess them.

Hermann is certainly right when she observes that von Neumann did not make the eigenvalues mistake imputed to him by Bell. But the rest of her reconstruction of von Neumann's argumentative strategy does not find support in von Neumann's text. As we have seen, von Neumann defines $\textbf{R} + \textbf{S}$ implicitly and in an operational way, such that \emph{by definition} it is true that $Exp(\textbf{R} + \textbf{S}) = Exp(\textbf{R}) + Exp(\textbf{S})$. He gives this definition \emph{without} invoking theoretical formalism, and only later makes the ``quantum assumption'' that all (already defined) physical quantities correspond in a one-to-one fashion to operators in Hilbert space. So Hermann's statement (see the above quotation) that according to von Neumann $\textbf{R} + \textbf{S}$ ``can be defined only as the quantity corresponding to the sum $R + S$ of the operators belonging to $\textbf{R}$ and $\textbf{S}$'' does not follow the line of von Neumann's argument. There seems to be a serious misunderstanding here on Hermann's part: indeed, there is no need at all for von Neumann to worry about a proof for the additivity of the expectation values of physical quantities, since in his argumentative set-up this is not an assumption at all but a conclusion  that follows analytically from the definition of $\textbf{R} + \textbf{S}$.

That von Neumann felt that he needed a proof for the ``additivity assumption'' and subsequently attempted to provide such a proof in several steps is therefore---strangely enough---a fabrication.\footnote{Remarkably, it seems that commentators have without exception accepted Hermann's account as an accurate description of von Neumann's argument, even though nothing can be found in von Neumann's text about the need for a proof of additivity.}

Hermann did not publish her 1933 manuscript, and her letter to Dirac appears to have remained unanswered. But she also sent her work to several other leading physicists and from her correspondence \cite{kay} it is clear that the essay was read by Bohr, Heisenberg and von Weizs\"{a}cker, who took it very seriously and discussed about a common reply.\footnote{In a letter of 17 December 1933 \cite{kay}, Gustav Heckmann, who was in close contact with all three of them, reported back from Copenhagen to Hermann: ``Sie nehmen Deine Arbeit voll und ganz \textit{ernst} und noch in den Tagen seines Hierseins wollte Heisenberg zusammen mit Bohr und einem Sch\"{u}ler Heisenbergs, Weizs\"{a}cker, die Antwort an Dich gemeinsam abfassen.'' } Von Weizs\"{a}cker took it on him to write\footnote{Weizs\"{a}cker to Hermann, 17 December 1933, in \cite{kay}.} to Hermann, thanking for the essay and responding to it on behalf of Bohr (who was too busy to write himself). Perhaps surprisingly, in his letter von Weizsa\"{a}cker immediately admits, without going into technical details, that one cannot prove that a deterministic extension of quantum mechanics is impossible by just looking at the mathematical formalism. Nevertheless, he maintains, there are weighty \emph{physical} grounds that together with the quantum formalism show that a return to determinism is most unlikely. He follows this up with a Bohrian analysis according to which the only physical quantities that should be considered are those that already occur in the language of classsical physics and that these cannot be imbedded in the quantum formalism without giving up determinism. He ends by mentioning that he will soon travel to Germany and would be happy to discuss these things in person. 

In a stenographic note to the letter Hermann observes with unmistakable surprise that von Weizs\"{a}cker does not object to her own conclusion that an impossibility proof for hidden variables cannot rely on the formalism all by itself but needs additional considerations.\footnote{Weisz\"{a}cker gibt zu: Der Indeterminismus folgt nur, wenn man die Voraussetzung macht, dass die beim heutigen Stand der Forschung ber\"{u}cksichtigten Observablen  alle Merkmale der atomaren Vorg\"{a}nge erfassen, und diese Voraussetzung l\"{a}sst sich aus dem Formalismus der Theorie nicht ableiten.}

The personal meeting with von Weizs\"{a}cker took place in Berlin on Friday January 12 1934, and on Sunday the 14th Hermann started writing a long letter (finished on the 17th)  in order to summarize points of agreement and disagreement. The first point that she mentions, and on which she asks von Weizs\"{a}cker's explicit confirmation in order to be absolutely sure, is that the mathematical formalism by itself does not exclude hidden variables\footnote{Der erste Punkt, in dem wir einig sind (wie mir bereits vor unserer Unterhaltung Ihr Brief zeigte), ist der, dass der mathematische Formalismus der Quantenmechanick nicht die Voraussetzungen f\"{u}r einen zwingenden Schluss auf den Indeterminismus enth\"{a}lt.}.  She adds, ``in this we are diametrically opposed to Born, Neumann and others''.\footnote{Wir stehen damit im Gegensatz zu Auffassungen von Born, Neumann u.a.} This addition again corroborates that Hermann incorrectly imputes to von Neumann the conclusion that no deterministic extension of quantum mechanics is possible at all, for purely mathematical reasons.  

In the rest of her letter of  January 14  Hermann argues about the validity of von Weizs\"{a}cker's and Bohr's additional physical arguments for the undesirability of hidden parameters. This is a point on which she and von Weizs\"{a}cker did not reach agreement, however. As becomes clear from several letters to her mother\footnote{\cite{kay}, letters of March 5, May 11, June 6 1934.}, Hermann now engaged in long and intensive personal discussions on this subject with Heisenberg. On June the 6th she writes to her mother that these discussions have finally led to a clear conclusion: Heisenberg was right all along, she was wrong, hidden variables and determinism in quantum mechanics are not physically acceptable. She continues however, in an elated tone, that this defeat has led her to a new and important insight.  Indeed, she has looked over and over again for a spot where Kant and Fries went wrong in their arguments for the validity of the law of causality, but could not find anything. Then it dawned upon her: their deductions were not in the least affected by the concessions she had to make to Heisenberg. It is true that she had to change her earlier position in some respects,  but it was possible to maintain the a priori validity of causality in its essence, in spite of the indeterminism of quantum mechanics.\footnote{Dann pl\"{o}tzlich eines Tages Einsicht: Diese Deduktion wird durch das, was ich Heisenberg zugestanden habe, \"{u}berhaupt nicht angetastet. Ich muss zwar meine Auffassung an einer ganz bestimmten Stelle \"{a}ndern; aber an einer Stelle, an der sich die \"{A}nderung in meiner philosophischen Ansicht einfach und verst\"{a}ndlich einf\"{u}gt. F\"{u}hre ich sie da konsequent durch, dann kann ich das Kausalgesetz in seinen mir wesentlichen St\"{u}cken aufrecht erhalten,... }         

These discussions and their upshot are reflected in a small treatise entitled \emph{The Natural-Philosophical Foundations of Quantum Mechanics}\footnote{\emph{Die naturphilosophischen Grundlagen der Quantenmechanik}} \cite{hermann1} that Hermann published in 1935. This treatise has two main messages. First, it is not possible to prove the completeness of quantum mechanics on purely mathematical grounds, and Hermann takes von Neumann to task for his supposed erroneous opinion to the contrary. Second, it \emph{is} possible to argue for the completeness of quantum mechanics on physical/philosophical grounds---but the resulting indeterminism should not be seen as conflicting with kantianism.

She starts, as in her 1933 manuscript, by reviewing and criticizing several arguments for indeterminism that can be found in the literature. In her chapter 7, about von Neumann, she takes over most of her earlier criticism of von Neumann's proof---although there are also some interesting differences, as we shall see. Hermann states \cite[pp.\@ 208--209]{crull}:
\begin{quote}
For the expectation value function $Exp(\textbf{R})$ thus defined by means of an ensemble of physical systems, which assigns a number to every physical quantity, Neumann assumes that $Exp(\textbf{R}+\textbf{S}) = Exp(\textbf{R})+Exp(\textbf{S})$. In words: the expectation value of a sum of physical quantities is equal to the sum of the expectation values of the two quantities. Neumann's proof stands or falls with this assumption.

For classical physics this assumption is trivial. So, too, it is for those quantum
mechanical features that do not mutually limit each other's measurability, thus between
which there are no uncertainty relations. Because for two such quantities, the
value of their sum is nothing other than the sum of the values that each of them
separately takes, from which follows immediately the same relation for the mean
values of these magnitudes. The relation is, however, not self-evident for quantum
mechanical quantities between which uncertainty relations hold, and in fact for the
reason that the sum of two such quantities is not immediately defined at all: since a
sharp measurement of one of them excludes that of the other, so that the two quantities
cannot simultaneously assume sharp values, the usual definition of the sum
of two quantities is not applicable. Only by the detour over certain mathematical
operators assigned to these quantities does the formalism introduce the concept of
a sum also for such quantities.
However, for the so-defined concept of the sum of two quantities that are not
simultaneously measurable, the formula given above requires a proof. Neumann carries
it out in two steps: since each ensemble of physical systems can be decomposed
into sub-ensembles whose elements agree with each other in terms of their wave
functions, then it follows, first, that the theorem in question needs to be proved only
for ensembles whose elements satisfy the condition of equal wave functions. But for
these ensembles Neumann relies on the fact that, in the context of the formalism, the
rule $((R+S)\varphi, \varphi) = (R\varphi, \varphi)+(S\varphi, \varphi)$ holds for the symbol $(R\varphi, \varphi)$, which represents
a number and is interpreted as the expectation value of the quantity $\textbf{R}$ in the state
$\varphi$. (Here $R$ and $S$ are the mathematical operators assigned to the quantities $\textbf{R}$ and
$\textbf{S}$; $\varphi$ specifies the wave function of the systems under consideration.) From this rule
Neumann concludes that for ensembles of systems with equal wave functions, and
therefore for all ensembles generally, the addition theorem for expectation values
holds also for quantities that are not simultaneously measurable.

\end{quote}

There are the same inaccuracies here as in the 1933 manuscript: 1. the additivity of the expectation values of physical quantities is represented as a substantive assumption made by von Neumann, whereas it is in fact a tautology; 2. it is suggested that von Neumann recognizes that he needs a proof for this assumption and attempts to gives this proof in two steps (being unaware that he is committing the fallacy of a \emph{petitio} while doing so). Hermann's verdict is indicated by the title of chapter 8 of her treatise, ``The Circle in Neumann's Proof''.

However, compared to the 1933 text, Hermann's 1935 recapitulation stays closer to von Neumann's original. In particular, it is clear that in the just-given quotation Hermann recognizes that ``von Neumann's additivity assumption'' is not formulated in terms of operators, but in terms of \emph{physical quantities}---even if she fails to notice the importance of the implicit definitions of some of these quantities. Related to this, in the above quotation Hermann seems to be more sensitive to the general line of von Neumann's proof, in which the essential step is the hypothesis that in quantum theories physical quantities should be represented by Hermitian operators. This is confirmed by what she writes after concluding that the proof is circular \cite[p.\@ 269]{crull}:
\begin{quote}
On the other hand, from the standpoint of Neumann's calculus one can argue
against this, that it is an axiomatic requirement that all physical
quantities are uniquely associated with certain Hermitian operators in a Hilbert
space, and that through the discovery of new features invalidating the present limits
of predictability, this association would inevitably be broken.
\end{quote}

This is exactly right! As we have seen, Von Neumann's proof does not rule out hidden variables in general, but does show that in hidden-variables theories the link between physical quantities and operators must be broken. Von Neumann concluded (as quoted in section \ref{proof}), that there presently is no reason to think that causality will be restored---the Hilbert space formalism is very well supported. But there is always the possibility that future data will necessitate a drastic revision of this verdict. A purely formal proof that no more complete theory than quantum mechanics will ever be possible cannot exist.\footnote{See also footnote \ref{Noteunderdet}.} Now compare this with Hermann's final conclusion of her chapter 8 \cite[p.\@ 270]{crull}:
\begin{quote}
By this consideration[\emph{i.e.\@ the previous quote}], however, the crucial physical question of whether the
progress of physical research can attain more precise predictions than are possible
today, cannot be twisted into the impossibly equivalent mathematical question of
whether such a development would be representable solely in terms of the quantum
mechanical operator calculus. There would need to be a compelling physical reason,
that not only the physical data known to date, but also all the results of research still
to be expected in the future are related to each other according to the axioms of
this formalism. But how should one find such a reason? The fact that the formalism
has so far proven itself, so that one is justified in seeing in it the appropriate
mathematical description of known natural connections, does not mean that the as
yet undiscovered natural law connections should also have the same mathematical
structure.
\end{quote}

This boils down to saying that the question of the adequacy of the Hilbert space formalism for the representation of physical reality is a physical rather than a mathematical issue. To attain certainty about this adequacy, also for the future, one should already know the outcomes of all future experiments. But this is impossible. So neither present-day physics nor mathematics can rule out the existence of hidden variables---future experience might make their introduction unavoidable.

Remarkably, in their final assessments of the situation Hermann and von Neumann substantially agree---even though Hermann apparently is unaware of this. Both think that hidden variables have only been excluded to the extent that they could fit in the Hilbert space formalism, and that it is an empirical question whether this means that they will never be needed. Von Neumann and Hermann differ in the way they formulate this conclusion: von Neumann maintains that given what we know about the remarkable history of quantum mechanics and its astounding successes, and the logical conflict between the structure of quantum mechanics and hidden variables, we have no good reasons to believe in a return of causality. Hermann, by contrast, emphasizes that it is impossible for physics, and even more so for mathematics, to definitively prove that such a return will never happen.

However, as we know from the outcome of her discussions with von Weizs\"{a}cker and Heisenberg, Grete Hermann does not intend to start a plea for hidden variables research. Her argument takes a very different turn: she contends that in spite of the possibility that future experiments will lead to a restoration of determinism, there is no need to be dissatisfied with quantum mechanics as it is, since it \emph{already} fulfils all requirements imposed by the law of causality. Therefore, even a neo-Kantian who considers causality as a necessary precondition of scientific knowledge may rest content. Hermann argues that in spite of the theory's indeterminism, quantum mechanics is causally complete because it specifies all factors that determine a measurement outcome. This is Hermann's new insight produced by her debates with Heisenberg.

This new analysis hinges on the interpretation Hermann now gives of the notion of causality. She acknowledges that quantum mechanics cannot \emph{predict} all results of measurements with certainty---but, she says, one should distinguish between predictability and causal determination. Indeed, once a measurement has taken place one \emph{can} tell a causal story about how that result has come into being, even if no such causal account was available beforehand. The key is that quantities that did not have definite values as judged before the measurement may possess such definite values from the viewpoint of the completed measurement. For example, an electron may be in a state without definite position, so that we are not able to predict what the outcome of a position measurement will be. But once a measurement has been performed, the electron does possess a well-defined position\footnote{In the ideal case that the electron has not been absorbed and not further disturbed.} and this makes it possible to construct with hindsight the path that it has followed.

Hermann's 1935 essay appeared in a relatively unknown philosophical series of neo-Kantian signature (\emph{Abhandlungen der Friesschen Schule}) and this is probably the reason that in the same year she published a 4-page summary in the well-known science journal \emph{Die Naturwissenschaften} \cite{hermann2}. In this summary Hermann does not mention von Neumann and his proof anymore, but takes it for granted that one can never completely rule out the existence of hidden parameters that in one way or another lift the quantum indeterminism: ``Whoever simply denies the possibility of such parameters, comes into conflict with the principle that experience is open.''\footnote{Wer die M\"{o}glichkeit solcher Merkmale slechthin leugnet, ger\"{a}t in Konflikt mit dem Satz von der Unabgeschlossenheit der Erfahrung.} Therefore, she says, there can only be one conclusive ground for abandoning as fundamentally fruitless the search for causes of an observed event: \emph{that one already knows these causes}.\footnote{Somit kann es nur einen einzigen hinreichenden Grund geben, das weitere Suchen nach den Ursachen eines beobachteten Vorganges als grunds\"{a}tzlich fruchtlos aufzugeben: \emph{den, da{\ss} man diese Ursachen bereits kennt}.} As we have seen a moment ago, according to Hermann this is exactly the situation that obtains in quantum mechanics. Even if an event was not predictable by quantum theory, it is possible to identify its necessary causes \emph{a posteriori}. Hermann concludes that ``it would therefore be futile to look for new physical parameters that have escaped discovery until now and that would be the causes of the event.''\footnote{Es w\"{a}re also sinnlos, f\"{u}r ihn [\emph{i.e.: den Vorgang}] in neuen, der Forschung bisher entgangenen physikalischen Merkmalen die Ursache seines Eintretens suchen zu wollen.}

To sum up her discussion of von Neumann's proof, in her 1933 manuscript Grete Hermann criticizes von Neumann for smuggling in the premise that only statistical collectives completely characterized by quantum mechanics need to be considered and for essentially proving on this basis that quantum mechanics cannot be made deterministic. In the treatise of 1935, Hermann's criticism is more sophisticated: she again makes the circularity claim, but she winds up by concluding that von Neumann only considered physical quantities that correspond to  quantum mechanical operators. As we have argued, in both cases details of her criticism, and the \emph{petitio} accusation, are mistaken. But Hermann is right in her conclusion, in the 1935 treatise, that von Neumann has only disproved the existence of hidden variables that can figure in theories with the same linear operators structure as quantum mechanics. This is not different from what von Neumann himself claimed, even if Hermann seems to think otherwise. Hermann's subsequent fervent philosophical defence of the causal completeness of quantum mechanics, and her appeal not to feel embarrassed by indeterminism and indefiniteness, bring her position effectively even closer to von Neumann's---that is, closer with respect to the question of completeness, not with respect to Hermann's neo-Kantian argumentation, of course.

\section{Later developments: Von Neumann and Bohm}\label{later}

Prior to the publication of Bohm's 1952 papers there seems to have been little in-depth discussion among leading physicists about concrete possibilities of hidden-variables theories \cite[pp.\@ 275--278]{jammer}. If anything, we encounter general points of view that are similar to that of von Neumann. For example, as Bacciagaluppi and Crull note \cite[p.\@ 139]{crull}, Pauli pointed out in a letter to Schr\"{o}dinger of 9 July 1935, commenting on the EPR paper, that the existence of dispersion-free states would lead to contradictions with quantum mechanics if it is assumed that the usual functional relations hold; Schr\"{o}dinger included this point in his paper on entanglement. The prominent philosopher of physics Hans Reichenbach declared in 1944 that although there are no logically compelling reasons for thinking that a return of determinism is impossible, there is no empirical indication that such a return is to be expected \cite[p.\@ 276]{jammer}. The Russian physicist Blokhinzev wrote, almost simultaneously with the appearance of Bohm's papers, that von Neumann's proof should not be seen as excluding hidden variables in all generality, and that a consistent hidden-variables theory will have to be cast in a form that deviates from the Hilbert space formalism of quantum mechanics \cite[p.\@ 277]{jammer}.

In this pre-war and early post-war literature there is hardly any reference to Grete Hermann's contributions, although we have seen that the Copenhagen circle knew her work, studied it, and was impressed by Hermann's cleverness. However, in contradistinction to what is often suggested in recent literature (\cite{mermin,seevinck}, \cite[pp.\@ 13--18]{herzenberg} are only a few examples), this lack of attention is not particularly surprising and does not call for external explanation. Indeed, Hermann published only two texts on our subject in the scholarly literature: \cite{hermann1} and \cite{hermann2}, respectively. The first appeared in a specialized neo-Kantian philosophy series and for that reason alone cannot be expected to have been widely read by physicists knowledgeable in quantum theory. Moreover, in this essay Hermann's conclusion with respect to the possibility of hidden variables does not strike one as spectacular: hidden variables are claimed to be possible in principle, but one has to go beyond the framework of quantum mechanics to accommodate them. This is what von Neumann himself had also concluded, and what seems the most reasonable position to adopt anyway (see also note \ref{Noteunderdet}). But even if one is not aware of von Neumann's conclusion and believes that he has definitively ruled out hidden variables on mathematical grounds, Hermann's argument that there are compelling \emph{philosophical} reasons for not engaging in a search for hidden variables does not seem too disturbing. In this context, the rather brief critical remarks in the 1935 essay about the details of von Neumann's proof may well appear a minor technical quibble.

Probably more directly relevant is that Hermann's second publication \cite{hermann2}, which had the aim of directing the attention of physicists to her more extensive philosophical treatise, does not mention von Neumann and his proof at all. It argues on philosophical grounds that there is no justification for considering quantum mechanics as incomplete and that one does not need to look for hidden variables---this hardly suggests an important novel point of view. It will not have helped that the philosophical line of argument of the paper does not come across as robust and convincing: it seems rather complacent to be content with causal links that can only be defined with hindsight, in situations without any predictability. As the philosopher Martin Strauss observed, in a rare contemporary reference to Hermann's work \cite[p.\@ 338]{strauss}, the causes that are identified in this \emph{post factum} way do not provide more information than the original quantum states, so that Hermann's scheme is ``empty''.\footnote{..., so da{\ss} das Kriterium hier nur erf\"{u}llt ist, weil es in dieser Form leer ist.} According to Jammer \cite[pp.\@ 209--210]{jammer}, Heisenberg at first responded positively to Hermann's ideas about causality, but had already changed his opinion in 1936.\footnote{Jammer himself diagnoses Hermann's claim of retrodictive causality as being unwarranted, since the \emph{post factum} causes have no explanatory significance \cite[p.\@ 209]{jammer}.} All in all, there is little reason to expect that Hermann's 1935 publications would have drawn much attention and would have impacted significantly on the physics community.

In 1952 David Bohm's two celebrated papers came out. In a report on his conversations with Bell, Jeremy Bernstein describes the effect of this event on Bell as follows \cite{bernstein}:
\begin{quote}
But early in 1952 something happened that changed everything. David Bohm published two papers which did just what von Neumann
had said was impossible. He created a hidden variable theory which was at least in some part deterministic and which reproduced all the results of the quantum theory. Something must have been terribly wrong with von Neumann.
\end{quote}

As can be expected, Bohm himself was also interested in the relation of his hidden-variables scheme to von Neumann's argument. He writes about von Neumann \cite[p.\@ 187]{bohm}: ``in his proof he has implicitly restricted himself to an excessively narrow class of hidden parameters and in this way has excluded from consideration precisely those types of hidden parameters which have been proposed in this paper.'' This characterization shows that Bohm was aware that von Neumann had only ruled out quite specific theoretical schemes. The qualification ``excessively narrow'' for the class of hidden variables for which the proof holds is of course subjective---from von Neumann's point of view this class (the one respecting Hilbert space structure) was the only one meriting serious attention; but for Bohm this was understandably different. Bohm made a serious study of von Neumann's proof and refers to it with page numbers. He gives the following more detailed diagnosis \cite[p.\@  187, see also note 9 on p.\@ 167]{bohm}:
\begin{quote}
  For example, if we consider two noncommuting observables, $p$ and $q$, then Von Neumann shows that it would be inconsistent with the usual rules of calculating quantum-mechanical probabilities to assume that there were in the observed system a set of hidden parameters which simultaneously determined the results of position and momentum ``observables.'' With this conclusion, we are in agreement. However, in our suggested new interpretation of the theory, the so-called ``observables'' are, as we have seen in Sec.\@5, not properties belonging to the observed system alone, but instead potentialities whose precise development depends just as much on the observing apparatus as on the observed system.
\end{quote}
The ``observables'' Bohm is here referring to are the experimental results actually found in measurements. What Bohm says is in agreement with the  conclusion of section \ref{proof}, namely that in hidden-variables theories experimental outcomes necessarily fail to always reflect the properties actually present in the system---this introduces an element of ``measurement contextuality'' in these theories.

What about von Neumann's own reaction to Bohm's papers? Bohm reported \cite{stoltzner}: ``It appears that von Neumann has agreed that my interpretation is logically consistent and leads to all results of the usual interpretation. (This I am told by some people.) Also, he came to a talk of mine and did not raise any objections.'' And on another occasion \cite[p.\@ 47]{freire}: ``von Neumann thinks my work correct, and even elegant, but he expects difficulties in extending it to spin.'' There is no indication that von Neumann felt that his earlier work had been refuted.

Bohm's articles rekindled the interest in hidden variables of Louis de Broglie. As de Broglie stated, in the meeting of the French Academy of Science of April 25th 1953 \cite[p.\@ 450]{debroglie}: ``Eighteen months ago, a young American physicist, David Bohm, has taken up (in a talented way I should add) my old ideas in the truncated and not well-defendable form of the pilot-wave.''\footnote{Il y a dix-huit mois, un jeune physicien am\'{e}ricain, M. David Bohm, a repris, d'ailleurs avec talent, mes anciennes id\'{e}es sous la forme tronqu\'{e}es et peu d\'{e}fendables de l'onde-pilote.} De Broglie explains that additional suggestions by Jean-Paul Vigier have since led him to a renewed investigation of his old hidden-variables theory, which seems to be worth-while after all (in spite of the objections at the 1927 Solvay conference). One of the things he has recently studied is the relation with von Neumann's theorem, and he has reached the following conclusion \cite[p.\@ 468]{debroglie}. Von Neumann assumed in his argumentation that the probability distributions for physical quantities, found in experiments, are defined by the states of the objects---but this is not so in his own (and Bohm's) theory. Admittedly, in these theories the \emph{positions} as defined by the states do correspond to what one measures, but this is \emph{not} so for measurements of momentum. Von Neumann's theorem is therefore not applicable to hidden-variables theories of the Bohm-de Broglie sort.\footnote{De Broglie confessed that he had not realized the limited domain of validity of the theorem before and had overrated its generality---``le th\'{e}or\`{e}me de M. von Neumann ne me para\^{i}t plus avoir la port\'{e}e que je lui attribuais moi-m\^{e}me dans ces derni\`{e}res ann\'{e}es.''}

Around 1952 both Bohm and de Broglie thus realized  that von Neumann's theorem pertained to hidden-variables theories of a specific kind, with a structure similar to that of quantum mechanics, in which measurement results correspond to physical quantities associated with the object (represented by operators in quantum mechanics). As it turns out, it was exactly the way in which the Bohm theory escaped von Neumann's argument, namely its introduction of contextuality, that provoked criticism from the leading contemporary participants in foundational debates. Indeed, although it is sometimes suggested that Bohm's theory was almost completely ignored, Einstein, Pauli and Heisenberg published papers in which they explained their reasons for not being convinced by Bohm's proposal \cite{myrvold}.

Einstein, following an earlier objection by Pauli, considered the case of a particle in an energy eigenstate, inside a box \cite[p.\@ 10]{myrvold}. As is well known, Bohm's theory says that the particle is at rest in the box; but when the box is opened and the particle's momentum measured, the usual non-zero results are found. These experimental values are therefore artifacts of the measurement, and the hidden reality of Bohm's world is completely different from what we have empirical access to. Pauli similarly pointed out that position and momentum are not treated symmetrically in Bohm's theory. Although a position measurement reveals a particle's actual position, Pauli states that \emph{Bohm eludes the von Neumann no-hidden-variables proof by contextualizing momentum} \cite[p.\@ 11]{myrvold}. Also Heisenberg objected to the same feature: ``for the measurements of position Bohm accepts the usual interpretation, for the measurements of velocity or momentum he rejects it'' \cite[p.\@ 13]{myrvold}.\footnote{As Myrvold \cite{myrvold} subsequently shows, the impossibility to assign non-contextual values to both position and momentum is not a peculiarity of the Bohm theory, but is a general feature of hidden-variables theories.}

\section{Conclusion}\label{conclusion}

In the history as we have reviewed it here, there is no mistake nor foolishness in von Neumann's proof. It is true that one may object to von Neumann's ``quantum assumption'' that all physical quantities should be theoretically represented by Hermitian operators in a Hilbert space, on the grounds that it is too strict. But von Neumann gives good inductive arguments for his assumption and is aware of its hypothetical character; it is certainly not a silly mistake to adopt a premise of this kind.

As we have also argued, Grete Hermann's claim of circularity of von Neumann's work was not justified. Nevertheless, her own 1935 assessment of what really followed from von Neumann's argument was not far from the mark---although her conclusion here was not spectacular (and stayed close to what von Neumann himself had claimed). Probably more important, her final verdict that it was unnecessary to look for hidden variables anyway seemed to be in accordance with well-known points of view. In addition, Hermann's purely philosophical arguments for this position, in her \emph{Die Naturwissenschaften} paper, come across as rather unphysical and shaky and will not have been able to excite many physicists. All in all, it is not surprising that Hermann's work did not cause a stir.

With the publication of Bohm's 1952 papers the discussion became more specific. Bohm himself analyzed in what respect his theory did not satisfy von Neumann's premises and was confirmed in his conclusions by de Broglie and Einstein, but also by the Copenhagenists Pauli and Heisenberg. Moreover, von Neumann himself saw no technical problems in Bohm's proposal. It is difficult to discern the contours of a concerted Copenhagen programme of indoctrination in this part of the history.

Bell recounts \cite{bell82} that he wondered about what von Neumann's argument had exactly proved when he saw Bohm's hidden-variables papers. He adds that he was unable to study the proof at the time, as no English translation of von Neumann's book was available, and that he therefore directed his attention to other subjects---although convinced that something had to be seriously wrong in von Neumann's arguments. It seems plausible that when he finally had a chance to read von Neumann's book, quite some years later, he looked for passages and sentences in which the expected error was clearly present. This would explain, for example, his emphasis on the last sentence of von Neumann's conclusion and his omission of the rest (section \ref{Bell}). It could also explain that he missed the precise structure of von Neumann's proof, in particular the role played in it by implicitly defined physical quantities.

What seems to be at least equally important is that Bell after 1966 made no secret of his foundational agenda. He fiercely defended the idea of hidden variables and made a firm stand against the Copenhagen hegemony. It goes without saying that for this agenda the exposure of a blatant error by one of the most prominent and revered (and in the meantime deceased) exponents of the ``powers that be'', an error allegedly used to justify the suppression of minority opinions, was an extremely effective rhetorical weapon. Indeed, the very prevalence of the standard story about how von Neumann's proof was shown up as a fake argument demonstrates this effectiveness.

It is possible that von Neumann's theorem has been invoked by some adherents of the Copenhagen interpretation as the final word against doubts about indeterminism and completeness. These must have been cases of ignorance, most likely mixed with a good dose of rhetoric. But the later disqualification of the theorem as a foolish fallacy by anti-Copenhagenists has certainly not been less ignorant and rhetorical.

\section*{Acknowledgment}
I am indebted to Guido Bacciagaluppi for alerting me to Hermann's correspondence about her 1933 essay (published in \cite{kay}), and also for giving me access to the final manuscript of \cite{crull} prior to its publication.

\end{document}